\documentclass[12pt]{article}  

\usepackage{graphicx,amsmath}
\usepackage{units}

\newlength{\dinwidth}
\newlength{\dinmargin}
\def\lapproxeq{\lower .7ex\hbox{$\;\stackrel{\textstyle
<}{\sim}\;$}}
\def\gapproxeq{\lower .7ex\hbox{$\;\stackrel{\textstyle
>}{\sim}\;$}}
\def\gtrsim{\lower .7ex\hbox{$\;\stackrel{\textstyle
>}{\sim}\;$}}
\def\lesim{\lower .7ex\hbox{$\;\stackrel{\textstyle
<}{\sim}\;$}}

\def\be{\begin{equation}}
\def\ee{\end{equation}}
\def\bea{\begin{eqnarray}}
\def\eea{\end{eqnarray}}

\def\GeV{\rm GeV}

\begin{document}

\begin{flushright}                                                    
IPPP/10/90  \\
DCPT/10/180 \\                                                    
\today \\                                                    
\end{flushright} 

\vspace*{0.5cm}

\begin{center}
{\Large \bf Towards a model which merges soft and hard high-energy $pp$ interactions\footnote{Talk presented by A.D. Martin at the Int. Workshop on Diffraction in High Energy Physics (Diffraction 2010), Otranto (Lecce), Italy, 10-15 Sept. 2010.}}
\vspace*{1cm}
                                                   
A.D. Martin$^a$, M.G. Ryskin$^{a,b}$ and V.A. Khoze$^{a,b}$ \\                                                    
                                                   
\vspace*{0.5cm}  
                                                                                                       
$^a$ Institute for Particle Physics Phenomenology, University of Durham, Durham, DH1 3LE \\                                                   
$^b$ Petersburg Nuclear Physics Institute, Gatchina, St.~Petersburg, 188300, Russia

\vspace*{0.5cm}                                                    
                                                    
\begin{abstract}                                                    
We seek a model which describes both the high-energy soft $pp$ data and has the perturbative QCD attributes expected in the low $x$, relatively low $Q^2$ domain. We describe the present status of this endeavour.

\end{abstract}                                                        
\vspace*{0.5cm}                                                    
                                                    
\end{center}


\section{Motivation}
Up to now there is no complete model which describes all facets of high-energy $pp$ interactions (elastic scattering, diffractive events, jet production etc.) on the same footing. We seek a model that not only describes pure soft high-energy low-$k_T$ data (via Pomeron exchange and Reggeon field theory), but which also extends into the large $k_T$ pQCD domain. Clearly to do this we shall need to introduce the {\it partonic} structure of the Pomeron.

What are the requirements of such a model? On the one hand, it should agree with the available soft high-energy $pp$ data, such as $\sigma_{\rm tot},~ d\sigma_{\rm el}/dt,~ d\sigma_{\rm SD}/dtdM^2,$ for $-t\lapproxeq 0.5 ~\GeV^2$. On the other hand, it should be in broad agreement with the known PDFs and diffractive PDFs for $x \lapproxeq 0.01$ and $Q^2 
\sim  4 ~\GeV^2$, as well as with the data for the single-particle inclusive $p_T$-distribution for $p^2_T \lapproxeq 4 ~\GeV^2$. Moreover, the model should satisfy $s$-channel unitarity. That is, the known large absorptive effects should be accounted for in terms of multi-Pomeron $t$-channel exchanges. Indeed, an important ingredient of a model seeking to link up the hard and soft regimes, is the extrapolation of the {\it partonic} (gluon-ladder) structure of the bare QCD Pomeron ({\it including} the gluon $k_T$ dependence along the ladder), into the soft regime taking account of these absorptive corrections, which become large as we go to smaller $k_T$.

\section{Soft and hard Pomerons?}
Often people speak of `soft' and `hard' Pomerons, so let us recall what is meant by these terms. The `soft' Pomeron is a vacuum-exchange object which drives soft high-energy interactions \cite{bkk}. It is not a simple Regge pole, but a non-local object. The rising $\sigma_{\rm tot}$ with energy means that multi-Pomeron diagrams (with Regge cuts) are necessary to restore unitarity. Total and elastic cross sections can be described, in the limited energy range up to Tevatron energies, by an effective pole with trajectory $\alpha_{\rm eff}\simeq 1.08+0.25t$ \cite{dl}, but this simple effective form breaks down at LHC energies. The `hard' or QCD Pomeron is described by the sum of ladder diagrams of Reggeized gluons with, in leading log$1/x$ BFKL \cite{fil}, a singularity which is a cut, and not a pole (or, with running $\alpha_s$ and boundary conditions at low $k_t$, a series of poles), in the complex angular momentum plane. When higher-order effects are included, the intercept of the `hard' Pomeron stabilizes to $\Delta=\alpha_P(0)-1 \simeq 0.3$.

From the discussion above, it is clear that our model will be based on the assumption that there exists only one Pomeron, which makes a smooth transition between the hard and soft regimes \cite{kmrrev}. What is the evidence that the soft Pomeron in the soft regime emerges from an extrapolation of the {\it bare} hard Pomeron? First, there is no irregularity in the HERA data in the transition region, $Q^2 \sim 0.3-2~ \GeV^2$. Second, a small slope, $\alpha' < 0.05~\GeV^{-2}$, of the {\it bare} Pomeron trajectory is found in the global analyses of the soft high-energy $pp$ data, after accounting for absorptive corrections and secondary Reggeons \cite{KMRnns1,um}. So, since $\alpha' \sim 1/k_T^2$, the typical values of $k_T$ inside the bare hard Pomeron amplitude are relatively large. Furthermore, these global analyses of soft high-energy data find that the intercept of the bare Pomeron trajectory is $\Delta=\alpha_P(0)-1 \simeq 0.3$, close to that of the QCD Pomeron. Moreover, HERA data on vector meson electroproduction show a power-like behaviour with energy which smoothly interpolates between the `effective' {\it soft} value $\alpha_P(0)\sim 1.1$ at $Q^2 \sim 0$, and the {\it hard} value $\sim 1.3$ at large $Q^2$. In summary, the bare  perturbative QCD Pomeron amplitude, with trajectory $\alpha_P \simeq 1.3+0t$, is subject to increasing absorptive effects as we go to smaller $k_T$, which allow it to yield the attributes of the soft Pomeron.

\section{Strategy}
We start with the ladder structure of the bare Pomeron amplitude, $\Omega_{ik}(y,{\bf k}_T,\bf{b})$. The $i,k$ subscripts denote the Good-Walker diffractive eigenstates, which allow for low-mass proton dissociation. The eigenstates are those combinations of $p,N^*,...$ which only undergo `elastic' scattering. Unitarity is imposed via a multichannel {\it eikonal}
\be
{\rm Im}T_{ik}~=~1-{\rm exp}(-\Omega_{ik}/2).
\ee
 The  bare amplitude, $\Omega$, satisfies an evolution equation in rapidity,
\be
\frac{\partial \Omega(y,{\bf k}_T,\bf{b})}{\partial y}~=~\int d^2k'_T~K({\bf k}_T,{\bf k}'_T)~\Omega(y,{\bf k}'_T,{\bf b}),
\ee
which generates the ladder structure by evolving from some input at $y=y_0$. At each step, ln$k_T^2$ and the impact parameter $\bf{b}$ can be changed, so, in principle, we have a three-variable integro-differential equation to solve. We use a simplified form of the kernel, $K$, which incorporates diffusion in ln$k_T^2$, and energy dependence $\Delta \sim 0.3$, as expected from BFKL. The $\bf{b}$ dependence during the evolution may be neglected, since it is proportional to the slope $\alpha'$, which is very small. Then the only $\bf{b}$ dependence comes from the input distribution.

The Multi-Pomeron contributions are included via absorptive factors of the form exp$(-\lambda\Omega_k/2)$, where $\lambda\Omega_k$ reflects the different opacity of the ``target'' $k$ felt by an intermediate parton, rather than the opacity $\Omega_k$ felt by the ``beam'' $i$. We expect $\lambda \sim 0.25$. If the rescattering involving intermediate partons is included (that is so-called {\it enhanced} rescattering), then the evolution up from $y=0$, or, to be precise, $y=y_0$ of the target, takes the form
\be
\frac{\partial \Omega_k(y,{\bf k}_T)}{\partial y}~=~\int d^2k'_T~{\rm exp}(-\lambda(\Omega_k(y)+\Omega_i(y'))/2)~K({\bf k}_T,{\bf k}'_T)~\Omega_k(y,{\bf k}'_T).
\label{eq:3}
\ee
Similarly the evolution down from the beam, $y'=Y-y$, is given by
\be
\frac{\partial \Omega_i(y',{\bf k}_T)}{\partial y}~=~\int d^2k'_T~{\rm exp}(-\lambda(\Omega_i(y')+\Omega_k(y))/2)~K({\bf k}_T,{\bf k}'_T)~\Omega_i(y',{\bf k}'_T).
\label{eq:4}
\ee
These two equations can be solved iteratively to give $\Omega_{ik}(y,{\bf k}_T,\bf{b})$, from which all observables can be calculated.

The aim is to study, in a semi-quantitative way,  the main features  of the soft and semi-hard interaction in terms of a realistic model with just a {\it few} physically-motivated parameters, and {\it not} to provide a many-parameter $\chi^2$-analysis of the data. In this way we hope that we can provide a better understanding of the physics which underlies the description of the data. Below we list a minimal set of parameters. There is basically one parameter (or sometimes two)  that is mainly responsible for each phenomena:
\begin{itemize}
\item The kernel $K$ is given in terms of $\Delta$ and $d$; $\Delta=\alpha_P(0)-1$ specifies the intercept of the {\it bare} Pomeron trajectory,  and $d$ controls the diffusion in ln$k_t$;
\item $\beta$  specifies the Pomeron-proton coupling;
\item $c_1$ and $c_2$ specify the proton radius and the corresponding form factor;
\item $\gamma_i$ specify the Good-Walker diffractive eigenstates, which are determined by low-mass diffractive
dissociation;
\item $\lambda$ which determines the strength of the triple- (and multi-) Pomeron couplings, which are constrained by data on high-mass diffractive dissociation;
\item $q_0$, the infrared cut-off, which together with $\beta $, controls the absolute value of the bare one-Pomeron exchange cross section;
\item $y_0$ which separates low- and high-mass diffraction.
\end{itemize}
Such an approach has been found to give a satisfactory description of soft high-energy $pp$ elastic and diffractive scattering data \cite{KMRnns1}. The absorptive effects are strong, and therefore we expect a relatively low $\sigma_{\rm tot}\sim 90$ mb at 14 TeV. However, here in (\ref{eq:3}) and (\ref{eq:4}), we include, for the first time, the ${\bf k}_T$ dependence of the opacity during the evolution, as well as the ${\bf b}$ dependence in the input distributions. We can therefore be more ambitious. We can now calculate the doubly-unintegrated gluon distribution. Integration over $\bf{b}$ and $k_T^2$ (up to $\mu^2$) then yields the gluon distribution $g(x,\mu^2)$ which is independently determined from global analyses of deep inelastic and related hard scattering data. Consistency between the two independent determinations for low $x$ and relatively low $Q^2$ is a tight constraint on the model. A similar comparison can be made for the diffractive gluon PDF. Moreover we can calculate the single-particle inclusive $p_T$-distribution and compare with data in the region $p_T^2 \lapproxeq 4~\GeV^2$.

\section{Discussion}
To achieve a simultaneous description of all of these soft and semi-hard phenomena is challenging. One ambiguity is that the form of the multi-Pomeron couplings is not known at present. The multi-Pomeron diagrams generated by the ``exp$(-\lambda\Omega/2)$'' absorptive factors in (\ref{eq:3}) and (\ref{eq:4}) correspond to $n \to m$ Pomeron couplings
\be
g^n_m~=~ nm~\lambda^{n+m-2}~g_N/2,~~~~~~~~{\rm for}~~~~n+m\ge 3,
\label{eq:5}
\ee
where $g_N$ is the Pomeron-proton coupling. These couplings are consistent with the conventional AGK cutting rules \cite{AGK} for the diagrams with triple-Pomeron vertices $(n+m=3)$; that is, for all the diagrams studied in the AGK paper, but are not consistent with the simplest generalisation of the AGK cutting rules for diagrams with vertices with $n+m >3$. Alternatively we could consider multi-Pomeron diagrams generated by absorptive factors ``$(1-{\rm exp}(-\lambda\Omega))/\lambda\Omega$'' in (\ref{eq:3}) and (\ref{eq:4}), which correspond to couplings of the form (\ref{eq:5}) but without the $nm$ factor. This leads to weaker absorption (with $\sigma_{\rm tot} \gapproxeq 100$ mb), but is consistent with the simplest extension of the AGK cutting rules. Hopefully, the constraints on the model will distinguish between these alternative $g^n_m$ forms.

The enhanced screening arising from both of these prescriptions is sufficient to restore unitarity during the evolution; eikonal screening gives just a little more absorption.  That is, the multi-Pomeron contributions are summed up in the absorptive factor which is included in {\it each} emission vertex. This is a very powerful result. It means that a relatively low number of new gluons will be produced during the evolution, which will greatly facilitate a Monte Carlo realisation of the model. 

A multi-Pomeron model has also been developed by Ostapchenko \cite{KPT}. It has a pure eikonal form for the multi-Pomeron couplings,
\be
g^n_m=(r_{3P}/4\pi)\gamma_P^{n+m-3},
\label{eq:gnmo}
\ee
with two parameters: $r_{3P}$ for the triple-Pomeron vertex and $\gamma_P$ to allow for the other vertices. In this case, for $r_{3P}/\gamma_P<\Delta$, the enhanced contribution inside a parton cascade is not strong enough to suppress the power growth of the bare Pomeron amplitude. Unitarity is only satisfied after eikonalization of the final amplitude. Unfortunately, the same $g^n_m$ are taken for the soft and hard components of the Pomeron. Thus the soft component screens the hard one, in contradiction to perturbative QCD. Moreover, the same PDFs are used for the proton-Pomeron and Pomeron-Pomeron interactions, which is probably why diffractive DIS is badly reproduced in Ref.~\cite{KPT}.

\end{document}